\documentclass[prl,twocolumn,showpacs,superscriptaddress]{revtex4}
\usepackage{graphicx}
\usepackage{amsmath}
\usepackage{bm}
\date{\today}
\begin{document}
\title{Domain-wall resistance in ferromagnetic (Ga,Mn)As} 
\author{D. Chiba}
\affiliation{ERATO Semiconductor Spintronics Project, Japan Science and Technology Agency, Japan}
\affiliation{Laboratory for Nanoelectronics and Spintronics, Research Institute of Electrical Communication, Tohoku University, Katahira 2-1-1, Aoba-ku, Sendai 980-8577, Japan}
\author{M. Yamanouchi}
\affiliation{Laboratory for Nanoelectronics and Spintronics, Research Institute of Electrical Communication, Tohoku University, Katahira 2-1-1, Aoba-ku, Sendai 980-8577, Japan}
\author{F. Matsukura}
\affiliation{Laboratory for Nanoelectronics and Spintronics, Research Institute of Electrical Communication, Tohoku University, Katahira 2-1-1, Aoba-ku, Sendai 980-8577, Japan}
\affiliation{ERATO Semiconductor Spintronics Project, Japan Science and Technology Agency, Japan}
\author{T. Dietl}
\affiliation{Institute of Physics, Polish Academy of Sciences, PL-02668 Warszawa, Poland; Institute of Theoretical Physics, Warsaw University, Poland}
\affiliation{Laboratory for Nanoelectronics and Spintronics, Research Institute of Electrical Communication, Tohoku University, Katahira 2-1-1, Aoba-ku, Sendai 980-8577, Japan}
\affiliation{ERATO Semiconductor Spintronics Project, Japan Science and Technology Agency, Japan}
\author{H. Ohno}
\affiliation{Laboratory for Nanoelectronics and Spintronics, Research Institute of Electrical Communication, Tohoku University, Katahira 2-1-1, Aoba-ku, Sendai 980-8577, Japan}
\affiliation{ERATO Semiconductor Spintronics Project, Japan Science and Technology Agency, Japan}
\email{ohno@riec.tohoku.ac.jp}

\begin{abstract}
A series of microstructures designed to pin domain-walls (DWs) in (Ga,Mn)As with perpendicular magnetic anisotropy has been employed to determine extrinsic and intrinsic contributions to DW resistance. The former is explained quantitatively as resulting from a polarity change in the Hall electric field at DW. The latter is one order of magnitude greater than a term brought about by anisotropic magnetoresistance and is shown to be consistent with disorder-induced misstracing of the carrier spins subject to spatially varying magnetization. 
\end{abstract}
\maketitle

The influence of domain-wall (DW) on transport properties of ferromagnetic materials, especially the sign and magnitude of the DW resistance (DWR), is attracting much attention because it is recognized that the description of spin flow across DW combines the physics of current-perpendicular-to-plane \cite{Greg96} and current-in-plane \cite{Levy97} giant magnetoresistances with the intricate question of spin dynamics in a spatially non-uniform field. Because transport of carriers in such a field gives rise to the Berry phase shift \cite{Ster92}, it was pointed out that DW may affect quantum localization phenomena \cite{Tata97,vaG99}. DWR is also related to the efficiency of DW displacement by spin-polarized current \cite{Tata04,Zhan04}; an issue not yet resolved despite the recent observations of the electrical DW motion in (Ga,Mn)As \cite{Yama04} and NiFe \cite{Yamag04, Verni04}. Determination of the magnitude and sign of DWR is thus an important and crucial step toward understanding the carrier and spin transport in a nonuniform magnetization background. 

In order to determine DWR experimentally, dense domain stripes \cite{Greg96,Vire96Dann02Marr04,Rued98} or geometrically confined patterns \cite{Ohta98,Tani99Lepa04Bunt05} have been employed in metallic systems, while in (Ga,Mn)As resistance changes associated with DW propagation between contacts have been monitored \cite{Tang04a}. Either increase \cite{Greg96,Vire96Dann02Marr04,Tani99Lepa04Bunt05} or decrease \cite{Rued98,Ohta98,Tang04a} of resistance by DW  has been found. The former is usually qualitatively interpreted in terms of spin-mistracing model put forward by Gregg {\it et al.}~\cite{Greg96} and developed further by Levy and Zhang \cite{Levy97}, whereas the negative DWR is assigned to the destruction of quantum localization by DW according to theory of Tatara and Fukuyama \cite{Tata97}. It is, however, not easy to measure directly small resistance changes associated with DW and to separate unambiguously and quantitatively various effects brought about by the presence of DW, such as anisotropic magnetoresistance (AMR) \cite{Rued98,Miya02} and the discontinuity of conductivity tensor components at DW \cite{Part74,Tang04b}. 

In this Letter, we report on the introduction of DWs at well-defined positions in microstructures of a (Ga,Mn)As layer with perpendicular magnetic easy axis. We combine magnetotransport and spatially resolved magneto-optical Kerr effect (MOKE) measurements to determine the DW characteristics. A series of microstructures with various geometries is used to separate extrinsic and intrinsic contributions to DWR. By finite-element computations we show that the extrinsic component originates from the sign change in the Hall electric field at DW. We then demonstrate that the intrinsic contribution can be explained quantitatively by Levy and Zhang theory \cite{Levy97}, developed within the s-d-type model directly applicable to the hole-mediated ferromagnetism in (Ga,Mn)As \cite{Diet01a}. Finally, we evaluate the AMR contribution and find that it is negligible under our experimental conditions.

The microstructures are fabricated from a thickness $t =$ 25~nm Ga$_{0.05}$Mn$_{0.95}$As layer grown by molecular beam epitaxy at 220$^{\circ\!}$C onto 500~nm In$_{0.15}$Ga$_{0.85}$As / 100 nm GaAs layers on a semi-insulating GaAs (001) substrate. The lattice-relaxed (In,Ga)As layer introduces tensile strain, which makes magnetic easy axis perpendicular to the film plane \cite{Diet01a}. A 180$^{\circ}$ DW is prepared at each boundary between non-etched and etched regions, defined by photolithography and wet etching as shown in Fig.~1(a), utilizing a slight dependence of Curie temperature $T_{\mathrm{C}}$, and thus coercivity $H_{\mathrm{c}}$, on $t$ \cite{Chib03Koed03,Yama04}. The channel is along [\={1}10] and the etched step is 7~nm high, which ensures a sufficient $H_{\mathrm{c}}$ contrast for a reproducible DW preparation. From the magnetic stiffness $A_{\mathrm{s}}$ and the uniaxial magnetic anisotropy energy density $K_{\mathrm{u}}$ corresponding to tensile strain and Mn concentration in question we evaluate the width of the Bloch wall to be $\delta_{\mathrm{W}} = \pi(A_{\mathrm{s}}/K_{\mathrm{u}})^{1/2} \approx 17$~nm \cite{Diet01b}. From the values of in-plane anisotropy energies determined by ferromagnetic-resonance spectra of similar films \cite{Yama05}, we find that the Bloch wall is energetically stabler than the N\'eel wall. We expect DWs to be at the etched side of the boundary, where the DW energy $4A_{\mathrm{d}}(A_{\mathrm{s}}K_{\mathrm{u}})^{1/2}$ is diminished owing to the reduced DW area $A_{\mathrm{d}}$, and smaller $A_{\mathrm{s}}$ and $K_{\mathrm{u}}$ due to lower $T_{\mathrm{C}}$.

\begin{figure}[t]
\includegraphics[width=3.3in]{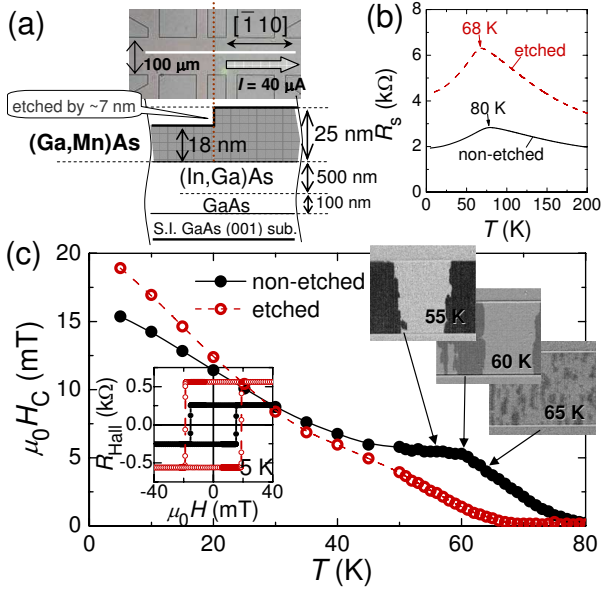}\vspace{-4mm}
\caption{[Color online] (a) Optical micrograph of Hall bar and schematic cross sectional view of structure under white line in the micrograph. Current ($I=40$~$\mu$A) direction is [\={1}10]. (b) Temperature dependence of sheet resistance $R_{\mathrm{s}}$ and (c) coercivity $\mu_0H_{\mathrm{c}}$ of non-etched and etched regions. Magnetic hysteresis loops measured by anomalous Hall effect at 5~K and three MOKE images at 55 - 65~K are displayed as insets to (c), where black (white) stripes in the channel correspond to negative (positive) values of $M$.} \label{fig:1}
\end{figure}

Properties of the non-etched and etched regions are measured by magnetotransport measurements as a function of temperature $T$ and magnetic field $H$ on the Hall bar shown in Fig.~1(a). The longitudinal and Hall resistivities provide the magnitude of the conductivity tensor components $\sigma_{ii}(T,H)$. The perpendicular easy axis makes it possible to determine magnetization $M(T,H)$ and, in particular, the coercivity $H_{\mathrm{c}}(T)$ by the Hall effect measurements as the Hall resistance is dominated by the anomalous Hall effect proportional to the perpendicular component of $M$. Figure 1(b) depicts temperature dependence of the sheet resistance $R_{\mathrm{s}}(T)$, whose {maxima are a measure of $T_{\mathrm{C}}$ of the non-etched and etched regions ($\sim\!80$~K and $\sim\!68$~K, respectively). These two regions exhibit also different temperature dependence of $H_{\mathrm{c}}(T)$, as shown in Fig.~1(c). The corresponding MOKE images displayed in Fig.~1(c) allow us to attribute the kink at 60~K in the $H_{\mathrm{c}}(T)$ dependence for the non-etched region to a boundary between DW nucleation (high temperatures) and DW propagation (low temperatures).

Figure 2(a) shows the magnetoresistance $\Delta R_{\mathrm{s}}(H) =R_{\mathrm{s}}(H)-R_{\mathrm{s}}(0)$ of the two regions for various orientations of the magnetic field $H(\theta,\phi)$, where $\theta$ and $\phi$ are the angles from [001] in $y$-$z$ plane and from [110] in $x$-$y$ plane, respectively, as depicted in Fig.~2(b). These data are relevant for the following two reasons. First, since the direction of magnetization changes across DW, information on the dependence of the resistivity on the angle between the magnetization and the current direction -- AMR -- is required to quantitatively assess its contribution \cite{Miya02}. Second, as shown in Fig.~2(c), resistance of (Ga,Mn)As exhibits a jump when magnetization reverses its direction at $H_{\mathrm{c}}$, which has to be taken into account when interpreting the origin of resistance changes caused by the formation of DW. 

\begin{figure}[t]
\includegraphics[width=3.3in]{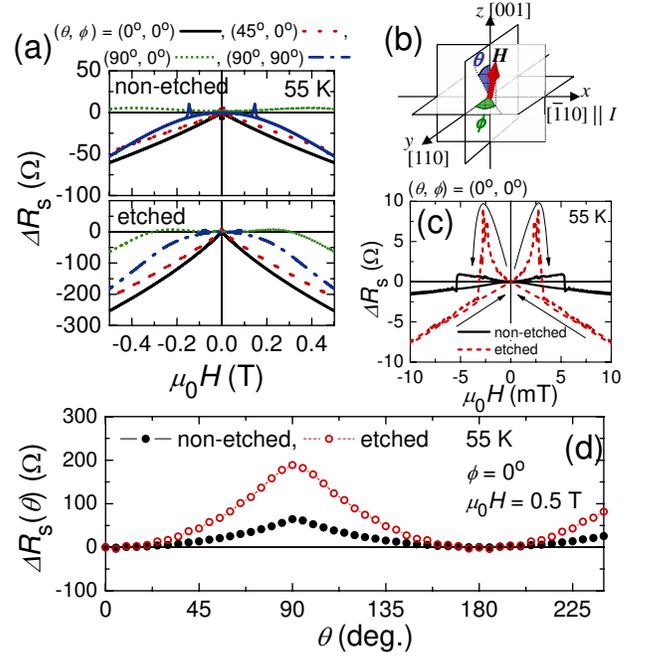}\vspace{-4mm}
\caption{[Color online] (a) Magnetoresistance $\Delta R_{\mathrm{s}}$ of the etched region for various directions of the magnetic field at 55~K. (b) Definition of relation between crystal orientation and angles $\theta$ and $\phi$ of the applied magnetic field. (c) Hysteresis cycle of $R_{\mathrm{s}}(H)$. (d) Dependence $\Delta R_{\mathrm{s}}(\theta)$ at 0.5~T.} \label{fig:2}
\end{figure}

For DWR measurements, we first fabricated two structures; one with the channel width $w = 100$~$\mu$m and length $l = 330$~$\mu$m [device A; Fig.~3(a)] and the other with the same dimension but with alternately etched (by $\sim\!7$~nm) and non-etched 30~$\mu$m long surface regions [device B; Fig.~3(b, c)]. Magnetoresistance and MOKE measurements are performed simultaneously at 45~K, as summarized in Fig.~4. When magnetic field is swept down after saturating magnetization in a sufficiently high positive field, magnetization of the etched regions (low $H_{\mathrm{c}}$), reverses gradually at $-3.0$\,-\,$-4.0$~mT in device B [Fig.~4(b)]. This is accompanied by resistance increases [see 1 and 2 in Fig.~4(b)], and then it reaches a maximum for a complete anti-parallel (AP) configuration (between $-4.0$ and $-5.8$~mT). As the field is further increased, magnetization of non-etched regions starts to reverse one-by-one ($-5.8$\,-\,$-10.9$~mT), resulting in a decrease of the number of DWs and staircase-like resistance reduction [3 to 9 in Fig.~4(b)]. The sum of individual resistance steps is $87$~$\Omega$ at 45~K in device B, which corrected for the DW independent resistance jump associated with the magnetization reversal, whose magnitude is linear with the magnetic field [see the curve of non-etched region in Fig.~2(c)]. This jump is $-46.4\mu_0H$[T] $\Omega$/DW, implying that the apparent DWR per DW is $+6.9$~$\Omega$. The complete AP configuration can be obtained with a good reproducibility in the temperature range 40\,-\,60~K. On the other hand, a smaller resistance increase is observed during DW propagation ($-5.2$\,-\,$-5.8$~mT) in device A [Fig.~4(a)]. 

\begin{figure}[t]
\includegraphics[width=3.3in]{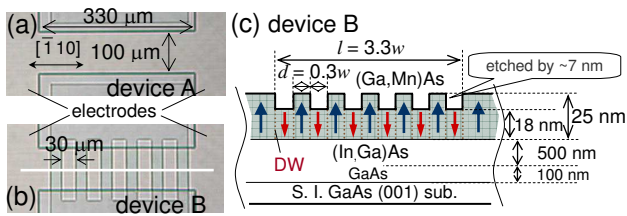}
\caption{[Color online] Optical micrographs of (a) unpatterned (device A) and (b) patterned (device B) layers for domain wall resistance measurements. (c) Schematic cross-sectional view under the white line of (b).} \label{fig:3}
\end{figure}

\begin{figure}[b]
\includegraphics[width=3.3in]{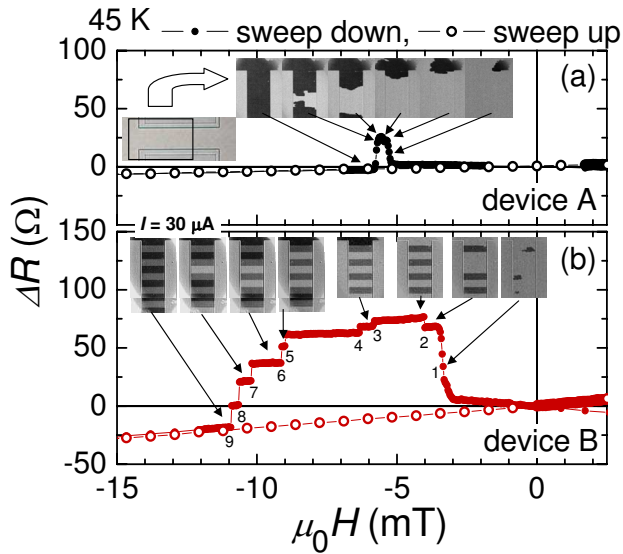}\vspace{-4mm}
\caption{[Color online] Magnetoresistance of device A (a) and device B (b) at 45~K (current $I=30$~$\mu$A). Insets show MOKE images in various magnetic fields disclosing relation between number of domain-walls and resistance.} \label{fig:4}
\end{figure} 

In the case of perpendicular easy axis, the polarity of the Hall electric field is opposite on the two sides of DW, leading to an extrinsic contribution to the apparent DWR \cite{Part74}. Using the conductivity tensors $\hat{\sigma}(T)$ obtained from the non-etched and etched Hall bars (Fig.~1), we evaluate this extrinsic contribution to DWR,  $R^{\mathrm{ext.}}$, from the continuity equation \cite{Part74,Tang04b}, $\mathrm{div}[\hat{\sigma}(x,y)\,\mathrm{grad}\, V(x,y)] =0$. Employing a standard finite element method, we obtain $R^{\mathrm{ext.}} = 6.3$~$\Omega$ per DW for a bar with 12 DW assuming $V$ = const.~at the bar ends at 45~K. The good agreement with the experimental value shows that the apparent DWR is dominated by this extrinsic effect. 

In order to extract the intrinsic DWR, we utilize the fact that the above contribution caused by the Hall effect is inversely proportional to the channel thickness $t$ and does not depend on the channel width $w$, whereas the intrinsic contribution is inversely proportional to the DW area $A_{\mathrm{d}} = wt_{\mathrm{etched}}$. To separate the two, we have fabricated microstructures with four different values of $w$, $w = 25$ (device C1), 50 (C2), 100 (C3), and 150 (C4)~$\mu$m, keeping the aspect ratios $l/w = 3.3$ and $d/w = 0.3$ the same, where $l$ is the total device length and $d$ is the distance between the etched (or non-etched) regions. Metal electrodes of 5~nm Cr/100~nm Au cover the two far end parts of the devices for ohmic contacts, leaving 6 etched and 5 non-etched regions in between. After saturating magnetization, the dispersion of resistance $R$ at $H = 0$ is below 10\% in devices C1--C4.  In the magnetic field, their resistance first increases, and then decreases in the staircase-like fashion displaying magnetization reversal of 5 non-etched islands (10 DWs) in all devices, as shown in Fig.~5(a). Solid lines have slopes expected from Fig.~2 for complete AP orientation of domains, and their extrapolation to $H=0$ gives DWR of each device. Thus determined DWR at 55~K is plotted as a function of inverse $w$ in Fig.~5(b), which clearly shows that DWR can be decomposed into a contribution independent of $w$ ($R^{\mathrm{ext.}}$) and linear in $w^{-1}$ (the intrinsic DWR, $R^{\mathrm{int.\!}}\!A_{\mathrm{d}}$). The values of $R^{\mathrm{ext.}}$ and $R^{\mathrm{int.\!}}\!A_{\mathrm{d}}$ are plotted as a function of temperature by open circles in Figs.~5(c) and 5(d), respectively. We again find that the finite-element results reproduce quantitatively the magnitude and temperature dependence of $R^{\mathrm{ext.}}$, as shown by open triangles in Fig.~5(c). The resistances of C1-C4 shows no systematic dependence on $w$; the largest difference being of the order of 10\% between samples C1 and C3. Such a spread in resistances is most probably caused by lateral wafer inhomogeneities and etching inaccuracy. We checked numerically that addition of DW independent interface resistance that increases the total resistance by 15\% ({\it i.e.}, greater than 10\% of $R$ dispersion) has no effect on the determined value of $R^{\mathrm{int.}}$. 

\begin{figure}[t]
\includegraphics[width=3.3in]{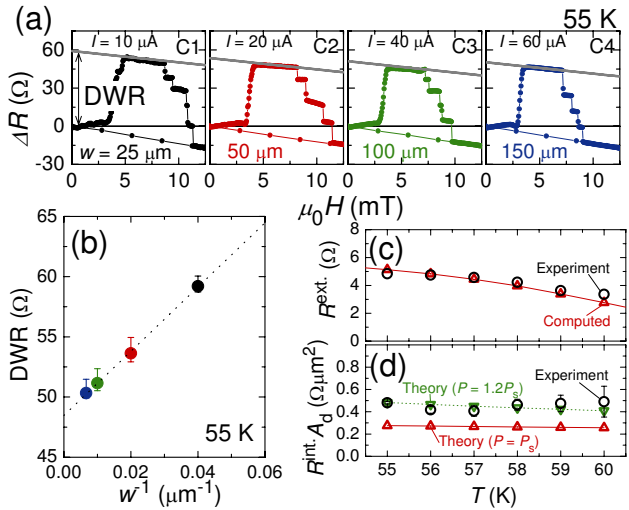}\vspace{-3mm}
\caption{[Color online] (a) Magnetoresistance of devices C1-C4 (current $I=10$, $20$, $40$, $60$~$\mu$A, respectively) with different channel width $w$. Solid lines have slopes expected from Fig.~2(c) for complete antiparallel orientation of domains, and their extrapolation to $H=0$ gives domain wall resistance (DWR). (b) DWR vs.~$w^{-1}$, which provides extrinsic and intrinsic contributions per one DW ($R^{\mathrm{ext.}}$ and $R^{\mathrm{int.\!}}\!A_{\mathrm{d}}$), shown by open circles in (c) and (d), respectively. Open triangles in (c) show computed $R^{\mathrm{ext.}}$. Open triangles and inverse triangles in (d) are theoretical calculated assuming that the conductance polarization is equal or 1.2 times greater that thermodynamic polarization, respectively.} \label{fig:5}
\end{figure} 

In order to elucidate the origin of the intrinsic DWR, we first calculate a contribution from AMR assuming Bloch DW, for which $\theta =\arccos[\tanh(\pi x/\delta_{\mathrm{W}})]$ and
\begin{equation}
R_{\mathrm{AMR}}A_{\mathrm{d}} = t\int_{-\infty}^{\infty}\mathrm{d}x\Delta R_{\mathrm{s}}[\theta(x)].
\end{equation}
By taking $\Delta R_{\mathrm{s}}(\theta)$ from Fig.~2, where $\Delta R_{\mathrm{s}}(\theta)$ is the difference between $\Delta R_{\mathrm{s}}$ at various $\theta$ and $\Delta R_{\mathrm{s}}$ at $\theta =0^{\circ}$, and $t$ suitable to the etched region we obtain $R_{\mathrm{AMR}}A_{\mathrm{d}} = 0.022$~$\Omega\,\mu$m$^2$ at 55~K, a factor of 20 smaller than the observed value of $R^{\mathrm{int.}}$. We then examine the effect of non-adiabatic contributions in carrier transport across DW. The hole precession time in the molecular field of the Mn spins, $\tau_{\mathrm{ex}} = \hbar g\mu_{\mathrm{B}}/| \beta | M$, is much shorter than the dwell time of the holes diffusing across DW, $\tau_{\mathrm{D}} = \delta_{\mathrm{W}}^2/D$, where the p-d exchange integral $\beta = -0.054$~eV\,nm$^3$ \cite{Diet01a} and $D$ is the diffusion constant.  Nevertheless, it has been known for some time in the Berry phase physics that disorder can introduce a considerable coupling between spin channels in the presence of a field texture \cite{Ster92}. This leads to a resistance enhancement \cite{Greg96,Levy97} of somewhat similar in origin to that known from theory of giant magnetoresistance in the current in-plane geometry. According to to theory of Levy and Zhang \cite{Levy97} for a simple parabolic s-type band, 
\begin{equation}
R^{\mathrm{int.}}A_{\mathrm{d}} = R_{\mathrm{s}}t\delta_{\mathrm{W}}\frac{4\xi^2P^2}{5(1-P^2)}\left[3+5(1-P^2)^{1/2}\right],
\end{equation}
where $\xi = \pi\hbar k_{\mathrm{F}}\tau_{\mathrm{ex}}/2m^*\delta_{\mathrm{W}}$ and the conductivity (current) polarization $P = (\sigma_{\uparrow} -\sigma_{\downarrow})/(\sigma_{\uparrow} + \sigma_{\downarrow})$. Since the heavy holes dominate we take $m^*=0.6m_0$ and $k_{\mathrm{F}}=1.8\times 10^9$~m$^{-1}$ that results from $p=2\times 10^{20}$~cm$^{-3}$ estimated from the p-d Zener model \cite{Diet01a}. For $P$ we adopt thermodynamic spin polarization $P_{\mathrm{s}}$ which away from the half-metallic situation is given by \cite{Diet01a}, $P_{\mathrm{s}}=6k_{\mathrm{B}}T_{\mathrm{C}}M/[(S+1)p\beta M_{\mathrm{S}}]$, where $M_{\mathrm{S}}$ is saturation magnetization. We see in Fig.~5(d) that theoretical predictions describe our experimental results quantitatively within a factor of two. If we assume that the complex valence band makes $P$ to be greater than $P_{\mathrm{s}}$ by a factor $1.2$, we find even better agreement in the whole temperature range studied, as shown by the dotted line. 

Finally, we note that previous study on (Ga,Mn)As reported a negative sign of DWR \cite{Tang04a}. There, (Ga,Mn)As with in-plane magnetic anisotropy was used, where AMR can give a negative contribution \cite{Jung02,Mats04}. In addition, measurements were carried out at lower temperatures (4.2~K) and on layers with resistivity by a factor of eight higher than the one used in the present study, making the destruction of quantum corrections that leads to negative DWR \cite{Tata97} more important than in our case.  

In summary, we have observed an increase of resistance associated with the DW formation in (Ga,Mn)As with perpendicular magnetic easy axis. This additional resistance contains a dominant contribution resulting from the alternating polarity of the Hall electric field at DWs. By carefully studying the geometry dependence of resistance, we have been able to extract an intrinsic component of DW resistance inversely proportional to the DW area. We relate the existence of this intrinsic term to a departure of spin transport across DW from the condition of adiabatic passage, a conclusion substantiated by a quantitative comparison to existing theory. We find the contribution of anisotropic magnetoresistance negligibly small under our experimental conditions. 

This work was supported in part by the IT Program of RR2002 from MEXT, Grant-in-Aids from MEXT/JSPS, Research Fellowship from JSPS, and the 21st Century COE program at Tohoku University.

\end{document}